# Numerical Simulations of 2-D Steady Free Convective flow with Heat and Mass Transfer in an Inclined Rectangular Domain

V.Ambethkar and D.Kushawaha

Department of Mathematics, Faculty of Mathematical Sciences

University of Delhi, Delhi, India.

Emails: vambethkar@maths.du.ac.in and vambethkar@gmail.com, dkushawaha@maths.du.ac.in and durgeshoct@gmail.com

## ABSTRACT

In this paper, we have used the QUICK scheme of the finite volume method to investigate the problem of steady 2-D free convective incompressible flow with heat and mass transfer in an inclined rectangular domain at different Rayleigh numbers in the range of $10^4 \leq Ra \leq 10^6$, for Prandtl number $Pr = 7.2$, and Lewis number $Le = 1$. We have used no-slip wall boundary conditions for the components of velocity and Neumann boundary conditions for temperature and concentration. We have used the QUICK scheme to discretize the governing equations along with the boundary conditions chosen in the present problem. The SIMPLE algorithm is adopted to compute the numerical solutions of flow variables, $U$-velocity, $V$-velocity, pressure, temperature, and concentration as well as the local and average Nusselt and Sherwood numbers at different Rayleigh numbers in the range mentioned above. Our numerical solutions for $U$-velocity along the vertical line through the geometric center has been compared with benchmark solutions available in the literature for $\lambda = 15^0,\ Ra = 10^4,\ Pr = 7.2,$ and $Le = 1$ and it has been found that it is in the good agreement. The present numerical results indicate that the $U$-velocity along the vertical line through the geometric center decreases with increasing the angles of inclinations and the $V$-velocity along the horizontal line through the geometric center doesn't vary with increasing the angles of inclinations. Pressure increases with increasing the angles of inclination. When the angles of inclination increase from $15^0$ to $45^0$, the intensity of streamlines decreases near the center and secondary cells are formed at the top and bottom of the center of the domain. As the Rayleigh number is increased from $10^4$ to $10^6$, the average Nusselt number is decreased for $\lambda = 15^0$, and decreases and then increases for $\lambda = 45^0$, whereas, the average Sherwood number increases for $\lambda = 15^0$ and decreases for $\lambda = 45^0$.

## 1. INTRODUCTION

The problem of 2-D steady free convective incompressible viscous fluid flow with heat and mass transfer has been the subject of intensive numerical computations in recent years. Fluid flows play an important role in various equipment and processes. Steady 2-D incompressible viscous flow coupled with heat and mass transfer in an inclined enclosure is a complex problem of high practical significance. This problem has received considerable attention due to its numerous engineering relevance in various disciplines, such as storage of radioactive nuclear waste materials, material transfer in ground water pollution, oil recovery processes, food processing, and the dispersion of chemical contaminants in various processes in the chemical industry. More often, fluid flow with heat and mass transfer are coupled in nature. Heat transfer is concerned with the physical process underlying the transport of thermal energy due to a temperature difference or gradient. All the process equipment used in engineering practice has to pass through a steady state in the beginning when the process is started, and they reach a steady-state after sufficient time has elapsed. Typical examples of steady heat transfer occur in heat exchangers, boiler tubes, cooling of cylinder heads in I.C. engines, heat treatment of engineering components and quenching of ingots, heating of electric irons, heating and cooling of buildings, freezing of foods, etc. Mass transfer is an important topic with broad industrial applications in mechanical, chemical and aerospace engineering. A few of the applications involving mass transfer are absorption and desorption, solvent extraction, evaporation of fuel in internal combustion engines, etc. Numerous everyday applications such as the dissolution of sugar in tea, drying of wood or clothes, evaporation of water vapor into the dry air, diffusion of smoke from a chimney into the atmosphere, etc. are also examples of mass diffusion. In many cases, it is interesting to note that heat and mass transfer occur simultaneously.

The boundary-layer regime for 2-D convection flow in a rectangular cavity with the two vertical walls has been studied by Gill [1]. Davis [2] studied steady laminar natural convection in an enclosed rectangular cavity. Newell and Schmidt [3] have studied laminar natural convection heat transfer within rectangular enclosures by using the Crank and Nicholson method. Natural convection in a shallow cavity with differentially heated end walls was studied by Cormack et al. [4]. Ozoe et al. [5] considered laminar natural convection in an inclined rectangular channel at various aspect ratios and angles with experimental measurements. Laminar natural convection heat transfer in a horizontal cavity with different end temperatures have been studied by Bejan and Tien [6]. Paolucci and Chenoweth [7] studied natural convection in shallow enclosures with differentially heated end walls. Numerical study on mode-transition of natural convection in differentially heated inclined enclosures was investigated by Soong et al. [8]. Tzeng et al. [9] studied the numerical investigation of transient flow-mode transition of laminar natural convection in an inclined enclosure. Interaction effects between surface radiation and turbulent natural convection in the square and rectangular enclosures were studied by Velusamy et al. [10]. Rahman and Sharif [11] investigated laminar natural convection in inclined rectangular enclosures of various aspect ratios. Snoussi et al. [12] numerically studied the natural convection flow resulting from the combined buoyancy effects of thermal and mass diffusion in a cavity with differentially heated side walls. Laminar natural convection in an inclined complicated cavity with spatially variable wall temperature was studied by Dalal and Das [13]. Dalal and Das [14] investigated natural convection in a rectangular cavity heated from below and uniformly cooled from the top and both sides. Basak et al. [15] studied effects of thermal boundary conditions on natural convection flows within a square cavity. Bilgen and Yedder [16] studied natural convection in an enclosure with heating and cooling by sinusoidal temperature profiles on one side. Deshmukh et al. [17] investigated natural circulation in cavities with uniform heat generation for different Prandtl number fluids. Basak et al. [18] investigated heatlines based natural convection analysis in tilted isosceles triangular enclosures with linearly heated inclined walls, and the effect of various orientations. Akiyama and Chong [19] studied natural convection with surface radiation in a square enclosure. Ambethkar and Kushawaha [20] have studied an unsteady 2-D incompressible flow with heat and mass transfer at low, moderate, and high Reynolds numbers in a rectangular domain using finite difference method. Ambethkar and Kushawaha [21] have investigated laminar incompressible flow and heat transfer in a four-sided lid-driven rectangular domain.

The above-mentioned literature survey pertinent to the present problem under consideration revealed that, to obtain highly accurate numerical solutions of the flow variables, we need to depend on a highly accurate and high resolution method like the SIMPLE algorithm of the finite volume approach. This algorithm is employed for computing unknown variables $U, V, P, T$, and $C$ simultaneously.

What motivated us is the enormous scope of applications of steady incompressible flow with heat and mass transfer in inclined domains as discussed earlier. The literature survey also revealed that the problem of fluid flow with heat and mass transfer in inclined rectangular domains, along with no-slip wall, temperature, and concentration boundary conditions, has not been widely studied numerically. Furthermore, to investigate the importance of the applications enumerated earlier, there is a need to determine numerical solutions of the unknown flow variables to fulfil this requirement, we present numerical simulations of the problem of fluid flow with heat and mass transfer in inclined rectangular domains, along with no-slip wall, temperature, and concentration boundary conditions, using the SIMPLE algorithm.

Our main target of this work is to numerically investigate fluid flow with heat and mass transfer in an inclined rectangular domain. We have used the QUICK scheme of finite volume methods to discretize the governing equations. The SIMPLE algorithm is adopted to compute the numerical solutions of the flow variables, $U$-velocity, $V$-velocity, $P$, $T$, $C$ and as well as the local and average Nusselt and Sherwood numbers for $\lambda = 15$ and 45 degrees, $10^4 \leq Ra \leq 10^6$, $Pr = 7.2, Le = 1$ and $N = 1$.

The summary of the layout of the current work is as follows: Section 2 describes the mathematical formulation that includes the physical description of the problem, governing equations, and boundary conditions. Section 3 describes the numerical solution of the governing equations along with a grid-independent test for velocity components and the local Nusselt and Sherwood numbers as well as the validity of results obtained with the benchmark solutions. Section 4 discusses the numerical results. Section 5 illustrates the conclusions of this study.

## 2. MATHEMATICAL FORMULATION

### 2.1 Physical description

The geometry of the problem considered in this work along with the boundary conditions is depicted in Figure 1. An inclined rectangular domain around the point (0.30, 0.15) in which laminar steady incompressible flow is considered. The horizontal walls have different constant temperatures and concentrations. The vertical walls are adiabatic and non-diffusive. The bottom wall is considered as the hot wall, and the top is as the cold wall.

We are assuming that, the values of $T$ on the hot and cold walls in such a way that the temperature defined on the bottom wall is greater than that of top wall. We are assuming the same convention for concentration. We are assumed that, at all four corner points of the computational domain, velocity components $(U, V)$ vanish. It may be noted here regarding in specifying the boundary conditions for pressure, the convention followed is that either the pressure at the boundary is given, or velocity components normal to the boundary are specified.

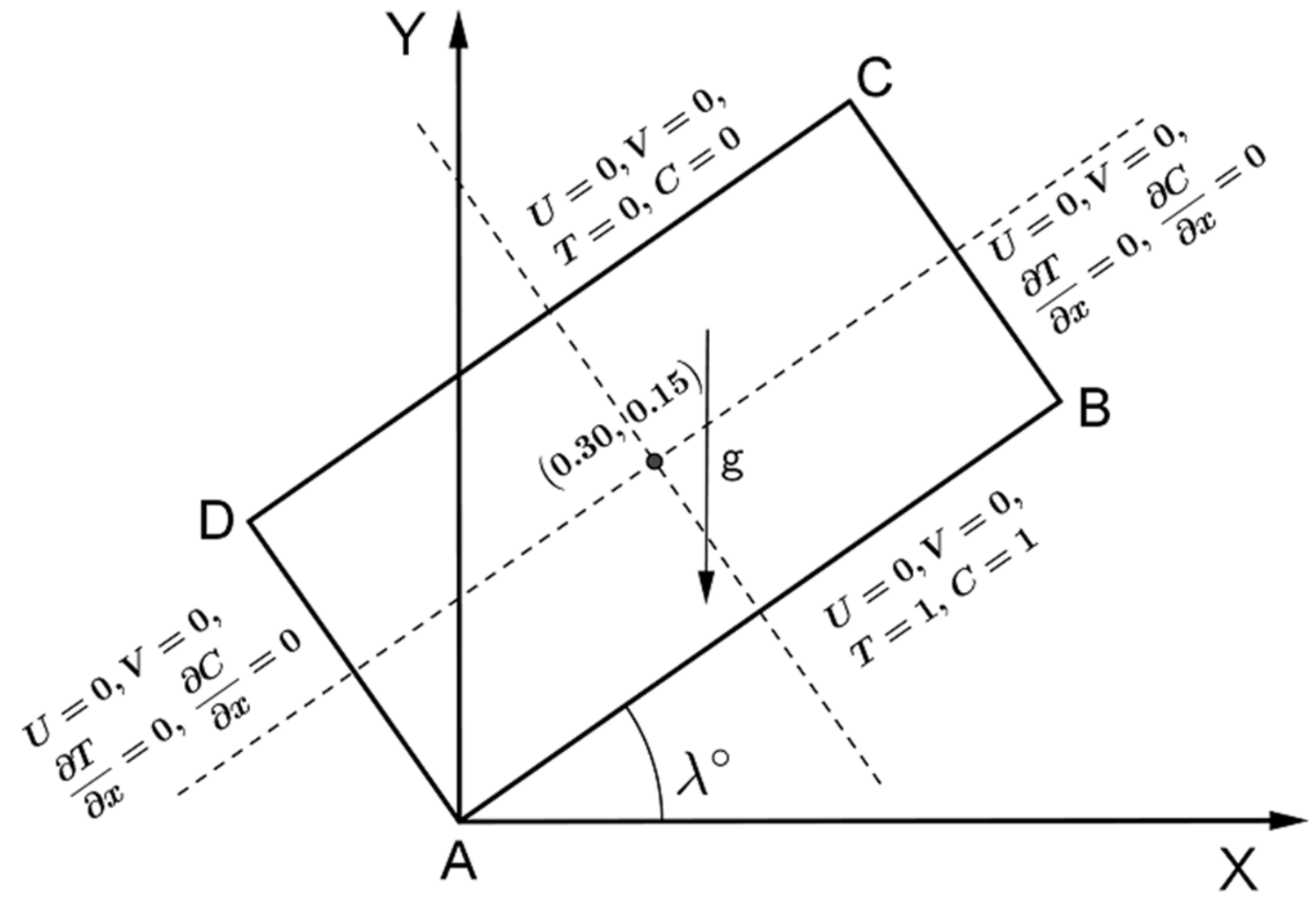

**Figure 1.** The geometry of the inclined rectangular domain problem.

## 2.2 Governing equations

The flow is assumed to be two-dimensional, steady-state, laminar, and the fluid is incompressible. The dimensionless forms of the governing equations are the continuity, the $X$- and $Y$-components of the Navier-Stokes equations, the energy equation, and the concentration equation, assuming negligible dissipation and constant thermo-physical properties, as given below:

Continuity: $$\frac{\partial U}{\partial X}+\frac{\partial V}{\partial Y}=0, \tag{1}$$

$X$-momentum: $$U\frac{\partial U}{\partial X}+V\frac{\partial U}{\partial Y}=-\frac{\partial P}{\partial X}+\Pr\left(\frac{\partial^2 U}{\partial X^2}+\frac{\partial^2 U}{\partial Y^2}\right)+Ra\Pr\left(T+NC\right)\sin\lambda, \tag{2}$$

$Y$-momentum: $$U\frac{\partial V}{\partial X}+V\frac{\partial V}{\partial Y}=-\frac{\partial P}{\partial Y}+\Pr\left(\frac{\partial^2 V}{\partial X^2}+\frac{\partial^2 V}{\partial Y^2}\right)+Ra\Pr\left(T+NC\right)\cos\lambda,$$

(3)

Energy equation: $U\frac{\partial T}{\partial X}+V\frac{\partial T}{\partial Y}=\frac{\partial^2 T}{\partial X^2}+\frac{\partial^2 T}{\partial Y^2},$ (4)

Mass transfer equation: $U\frac{\partial C}{\partial X}+V\frac{\partial C}{\partial Y}=\frac{1}{Le}\left(\frac{\partial^2 C}{\partial X^2}+\frac{\partial^2 C}{\partial Y^2}\right),$ (5) where $U, V, P, T, C, Pr, Ra, Le,$ and $N$ are the dimensionless velocity components in $X$-and $Y$-directions, the dimensionless pressure, the dimensionless temperature, the dimensionless concentration, the Prandtl number, the Rayleigh number, the Lewis number, and the buoyancy ratio respectively. We define the following non-dimensional variables.

$$\left.\begin{aligned} &X=\frac{x}{L},\ Y=\frac{y}{L},\ U=\frac{uL}{\alpha},\ V=\frac{vL}{\alpha},\ T=\frac{T'-T'_c}{T'_h-T'_c},\\ &C=\frac{C'-C'_c}{C'_h-C'_c},\ P=\frac{pL^3}{\rho\alpha^2},\ \Pr=\frac{\nu}{\alpha},\ Le=\frac{\alpha}{D},\\ &Ra=\frac{g\beta_{T'}(T'_h-T'_c)L^3\Pr}{\nu^2},\ N=\frac{\beta_{C'}(C'_h-C'_c)}{\beta_{T'}(T'_h-T'_c)}.\end{aligned}\right\} \quad (6)$$

The no-slip wall boundary conditions are given by:

$$\begin{aligned} &U(X,Y)=0, V(X,Y)=0, T=C=1, \text{ along wall AB},\\ &U(X,Y)=0, V(X,Y)=0, \frac{\partial T}{\partial X}=\frac{\partial C}{\partial X}=0, \text{ along wall BC},\\ &U(X,Y)=0, V(X,Y)=0, T=C=0, \text{ along wall CD},\\ &U(X,Y)=0, V(X,Y)=0, \frac{\partial T}{\partial X}=\frac{\partial C}{\partial X}=0, \text{ along wall DA}.\end{aligned} \quad (7)$$

The stream function is calculated from the definition

$$U=\frac{\partial\psi}{\partial Y} \text{ and } V=-\frac{\partial\psi}{\partial X}. \quad (8)$$

Which yields a single equation:

$$\frac{\partial^2\psi}{\partial X^2}+\frac{\partial^2\psi}{\partial Y^2}=\frac{\partial U}{\partial Y}-\frac{\partial V}{\partial X}. \quad (9)$$

The sign convention is as follows: a positive sign of $\psi$ denotes anti-clockwise circulation and clockwise circulation is represented by negative sign of $\psi$. It is taken that $\psi = 0$ at the solid boundaries.

The heat transfer coefficient in terms of the local Nusselt number $(Nu)$ is defined by

$$Nu_X = -\frac{\partial T}{\partial Y}, \tag{10}$$

The total heat transfer in terms of the average Nusselt is defined by

$$\overline{Nu} = \frac{1}{AR}\int_0^{0.6} Nu_X \, dY. \tag{11}$$

The mass transfer coefficient in terms of the local Sherwood number $(Sh)$ is defined by

$$Sh_X = -\frac{\partial C}{\partial Y}, \tag{12}$$

The total heat transfer in terms of the average Sherwood number is given by

$$\overline{Sh} = \frac{1}{AR}\int_0^{0.6} Sh_X \, dY. \tag{13}$$

## 3. VALIDATION OF THE NUMERICAL SOLUTIONS

We have used a finite volume method to solve the governing equations after reduced into discretizing equations by the QUICK scheme. The SIMPLE algorithm is adopted in solving the discretized momentum, energy, mass transfer, and the pressure correction equations. Discretization of the governing equations using the QUICK scheme is being skipped here as it is available in the literature [22].

All the results obtained in this study converged to a maximum residual of $10^{-6}$. Furthermore, we considered three different grid systems: $41 \times 21, 61 \times 31,$ and $81 \times 41$ for $\lambda = 15^0, Ra = 10^4, pr = 7.2, Le = 1,$ and $N = 1$ to ensure grid-independent results and $61 \times 31$ grid points were considered to obtain an accurate solution in the entire computation of this study.

To check the validity of our present computer code used to obtain the numerical result of $U$-velocity, we have compared our present result with those benchmark result is given by Paolucci and Chenoweth [7] and it has been found that it is in the good agreement.

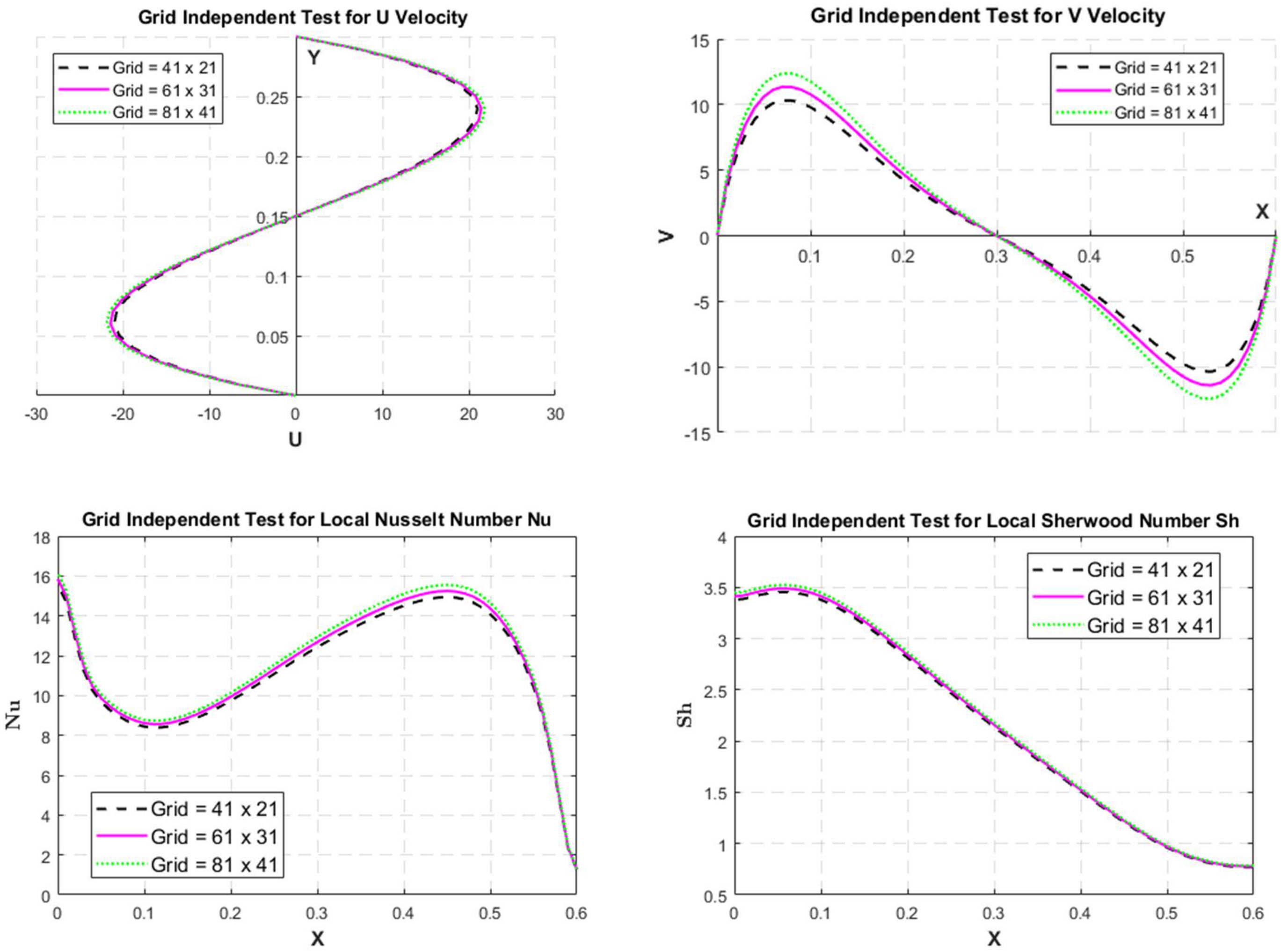

**Figure 2.** Grid independence test for $U$-velocity, $V$-velocity, local Nusselt and Sherwood numbers for $\lambda = 15^0, Ra = 10^4, Pr = 7.2, Le = 1,$ and $N = 1$.

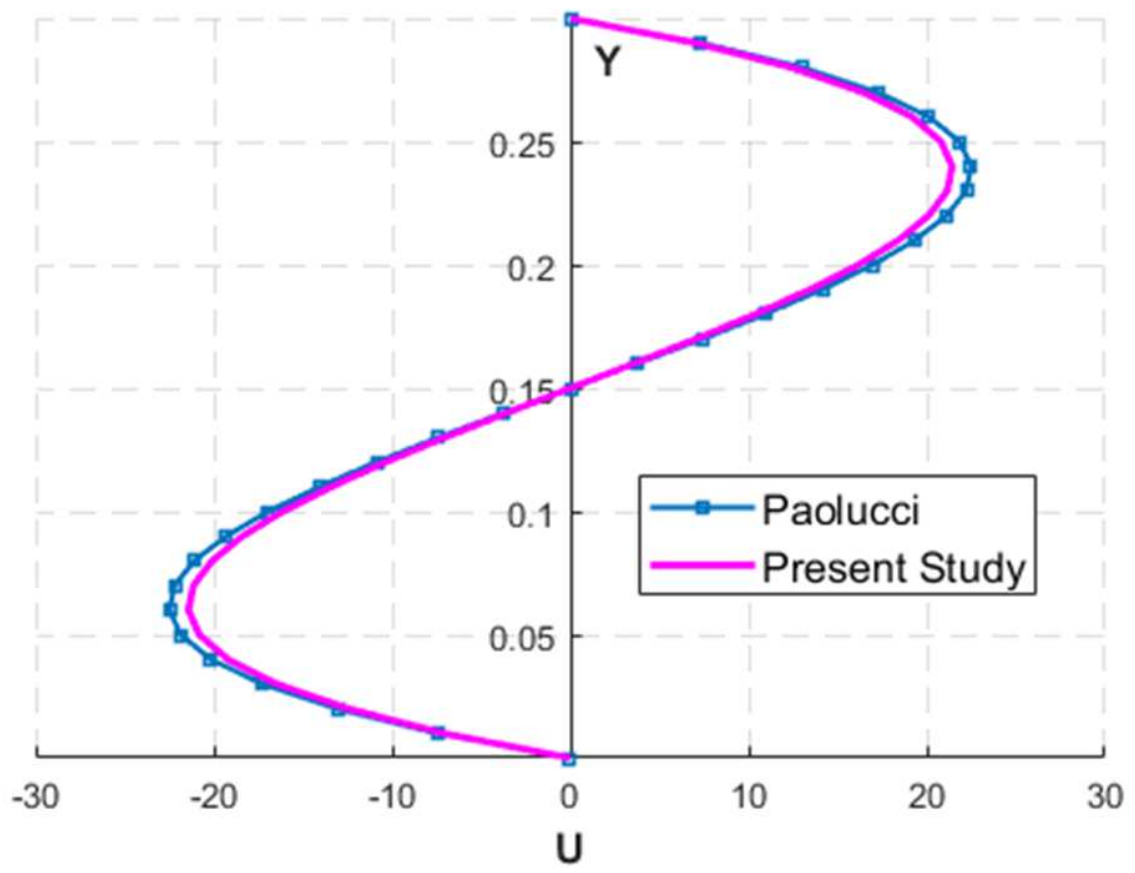

**Figure 3.** Comparison of the numerical results of $U$-velocity along the vertical line through the geometric center of the domain.

## 4. RESULTS AND DISCUSSION

Fluid flow, temperature, and concentration fields in inclined rectangular domains are examined. The numerical simulations are performed for the different Rayleigh numbers in the range of $10^4 \leq Ra \leq 10^6$, different angle of inclination namely, $\lambda = 15^0$ and $45^0$, Prandlt number $Pr = 7.2$, Lewis number $Le = 1$, and buoyancy ratio $N = 1$. We present the results here including the pressure contours, streamlines, isotherms, iso-concentrations, $U$-velocity contours, $V$-velocity contours, $U$-velocity profiles along the vertical line through the geometric center, $V$-velocity profiles along the horizontal line through the geometric center, the local Nusselt and Sherwood numbers along the cold wall of the domain, and the average Nusselt and Sherwood numbers for above mentioned parameters.

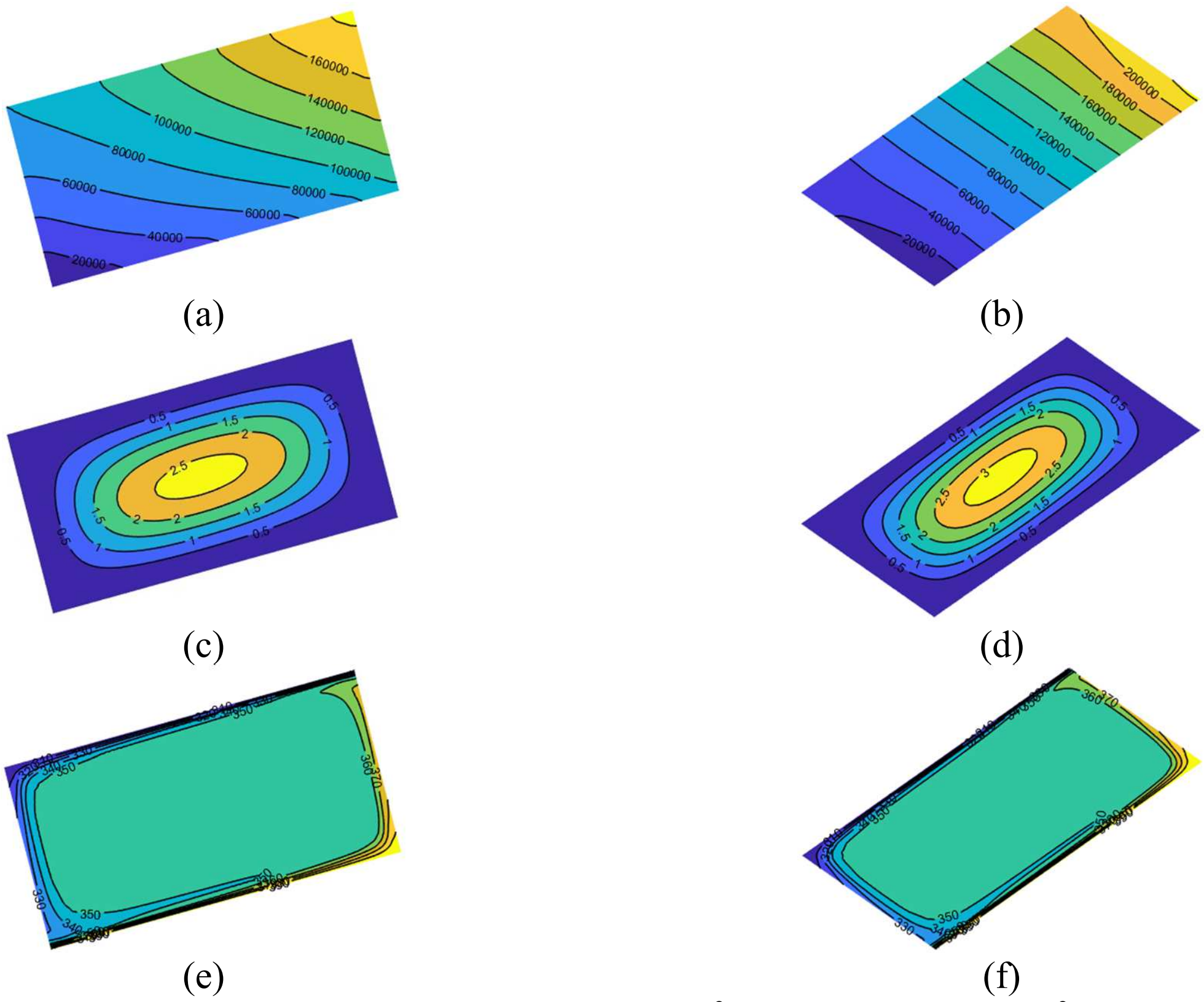

**Figure 4.** Pressure contours, streamlines, and isotherms for $\lambda = 15^0$ (left column) and $\lambda = 45^0$ (right column), $Ra = 10^4, Pr = 7.2, Le = 1,$ and $N = 1$.

From Figures 4(a) and (b), it is observed that pressure is low at the bottom left corner and high at the top right corner of the domain for $\lambda = 15^0$ and $\lambda = 45^0$. When the angle of inclination increases from $15^0$ to $45^0$, the numerical values of pressure increase but the nature of the contours are same. From figures 4(c) and (d), we can see that an anti-clockwise circulation is formed at the center of the domain, and the intensity of the streamlines are strong near the center and weaker near to the boundaries of the domain. When the angle of inclination increases from $15^0$ to $45^0$, the intensity increases at the center and decreases near the boundaries but the nature of the streamlines are same. From figures 4(e) and (f), it is observed that from the bottom to top temperature decreases and the same thing happens from left to right. When the angle of inclination increases from $15^0$ to $45^0$, similar patterns are found for isotherms.

From Figures 5(a) and (b), the iso-concentration contours show that concentration decreases from the bottom to top. When the angle of inclination increases from $15^0$ to $45^0$, iso-concentrations follow the same pattern. From figure 5(c) we observed that there are two circulations; one is anti-clockwise and another one, clockwise, are formed in the domain for $\lambda = 15^0$. An anti-clockwise one is formed at the bottom, and the intensity is high near the center of the circulation; a clockwise one is formed at the top, and the intensity is weaker near the center of the circulation. From figure 5(d) we observed that there are two circulations; one is anti-clockwise and the other, clockwise, are formed in the domain for $\lambda = 45^0$. The anti-clockwise one is formed at the bottom left, and the intensity is high near the center of the circulation; the clockwise one is formed at the top right, and the intensity is weaker near the center of the circulation. When the angle of inclination increases from $\lambda = 15^0$ to $45^0$, the intensity decreases at the center of the bottom circulation and increases at the center of the top circulation. From figures 5(e) and (f) we observed that there are two circulations, one is anti-clockwise and other one, clockwise, are formed in the domain for $\lambda = 15^0$ and $\lambda = 45^0$. The anti-clockwise one is formed at the bottom right corner, and the intensity is high near the center of the circulation; the clockwise one is formed at the top left corner, and the intensity is weaker near the center of the circulation. When the angle of inclination increases from $\lambda = 15^0$ to $45^0$, the intensity increases at the center of the bottom right corner circulation and decreases at the center of the top left corner circulation.

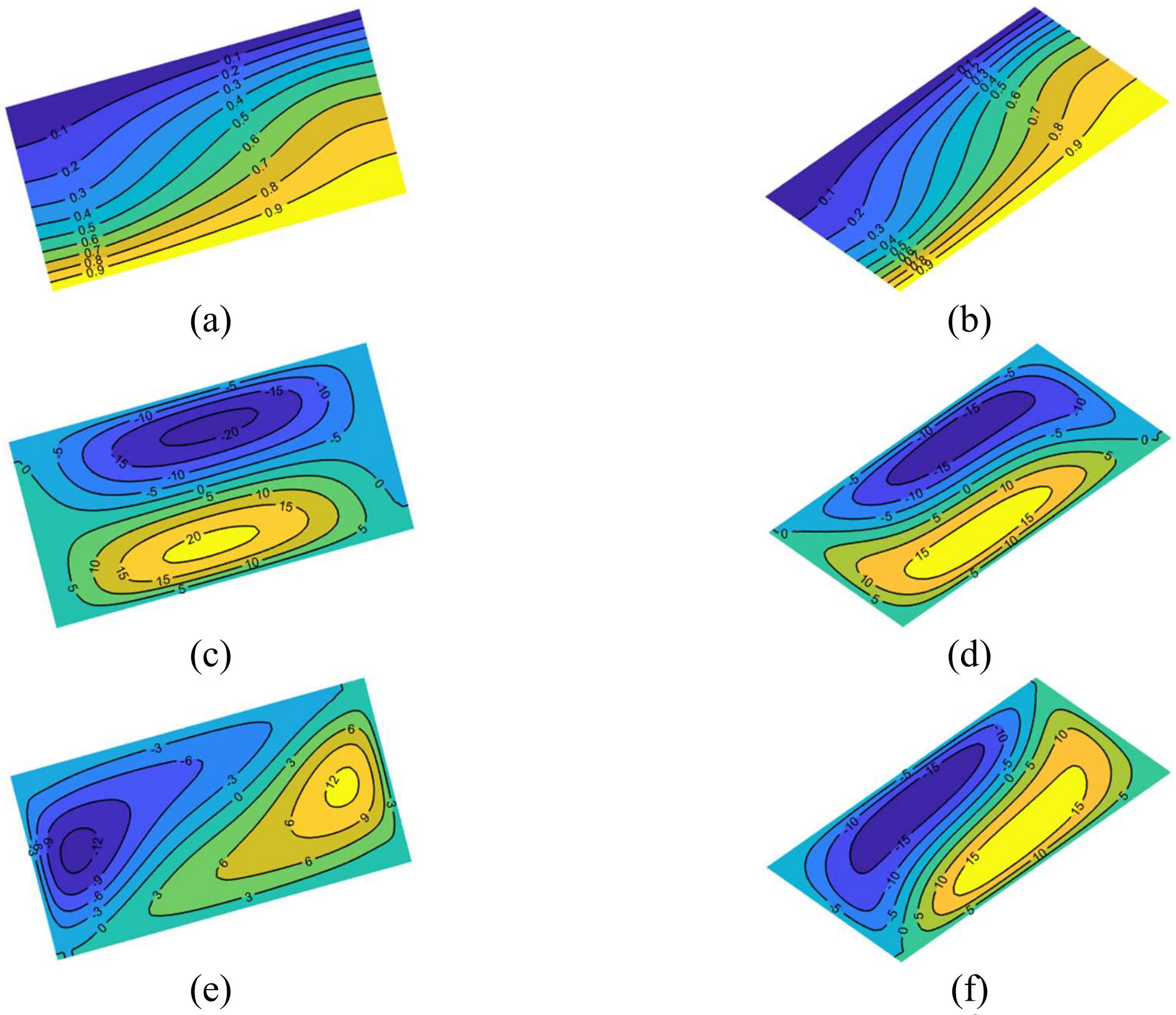

**Figure 5.** Iso-concentrations, $U$-velocity contours, and $V$-velocity contours for $\lambda = 15^0$ (left column) and $\lambda = 45^0$ (right column), $Ra = 10^4, Pr = 7.2, Le = 1,$ and $N = 1$.

From Figures 6(a) and (b), it is observed that pressure is low at the bottom left corner and high at the top right corner of the domain for $\lambda = 15^0$ and $\lambda = 45^0$. When the angle of inclination increases from $15^0$ to $45^0$, the numerical values of pressure increase but the nature of the contours are same. From figures 6(c) and (d) there is an anti-clockwise circulation formed at the center of the domain, and the intensity of the streamlines are strong near the center and weaker near the boundaries of the domain. When the angle of inclination increases from $15^0$ to $45^0$, the intensity increases at the center and decreases near the boundaries but the nature of the streamlines are same. From figures 6(e) and (f) we can see that from the bottom to top temperature decreases and the same thing happens from left to right. When the angle of inclination increases from $15^0$ to $45^0$, similar patterns are found for isotherms.

From Figures 7(a) and (b), the iso-concentration contours show that concentration decreases from bottom to top. When the angle of inclination increases from $15^0$ to $45^0$, iso-concentrations follow the same pattern. From figure 7(c) we observed that there are two circulations, one is anti-clockwise and another one, clockwise, are formed in the domain for $\lambda = 15^0$. The anti-clockwise one is formed at the bottom, and the intensity is high near the center of the circulation; the clockwise one is formed at the top, and the intensity is weaker near the center of the circulation. From figure 7(d) we observed that there are two circulations, one is anti-clockwise and another, clockwise, are formed in the domain for $\lambda = 45^0$. The anti-clockwise one is formed at the bottom left, and the intensity is high near the center of the circulation; the clockwise one is formed at the top right, and the intensity is weaker near the center of the circulation. When the angle of inclination increases from $\lambda = 15^0$ to $45^0$, the intensity decreases at the center of the bottom circulation and increases at the center of the top circulation. From figures 7(e) and (f) we observed that there are two circulations, one is anti-clockwise and other one, clockwise, are formed in the domain for $\lambda = 15^0$ and $\lambda = 45^0$. The anti-clockwise one is formed at the bottom right corner, and the intensity is high near the center of the circulation; the clockwise one is formed at the top left corner, and the intensity is weaker near the center of the circulation. When the angle of inclination increases from $\lambda = 15^0$ to $45^0$, the intensity decreases at the center of the bottom right corner circulation and increases at the center of the top left corner circulation.

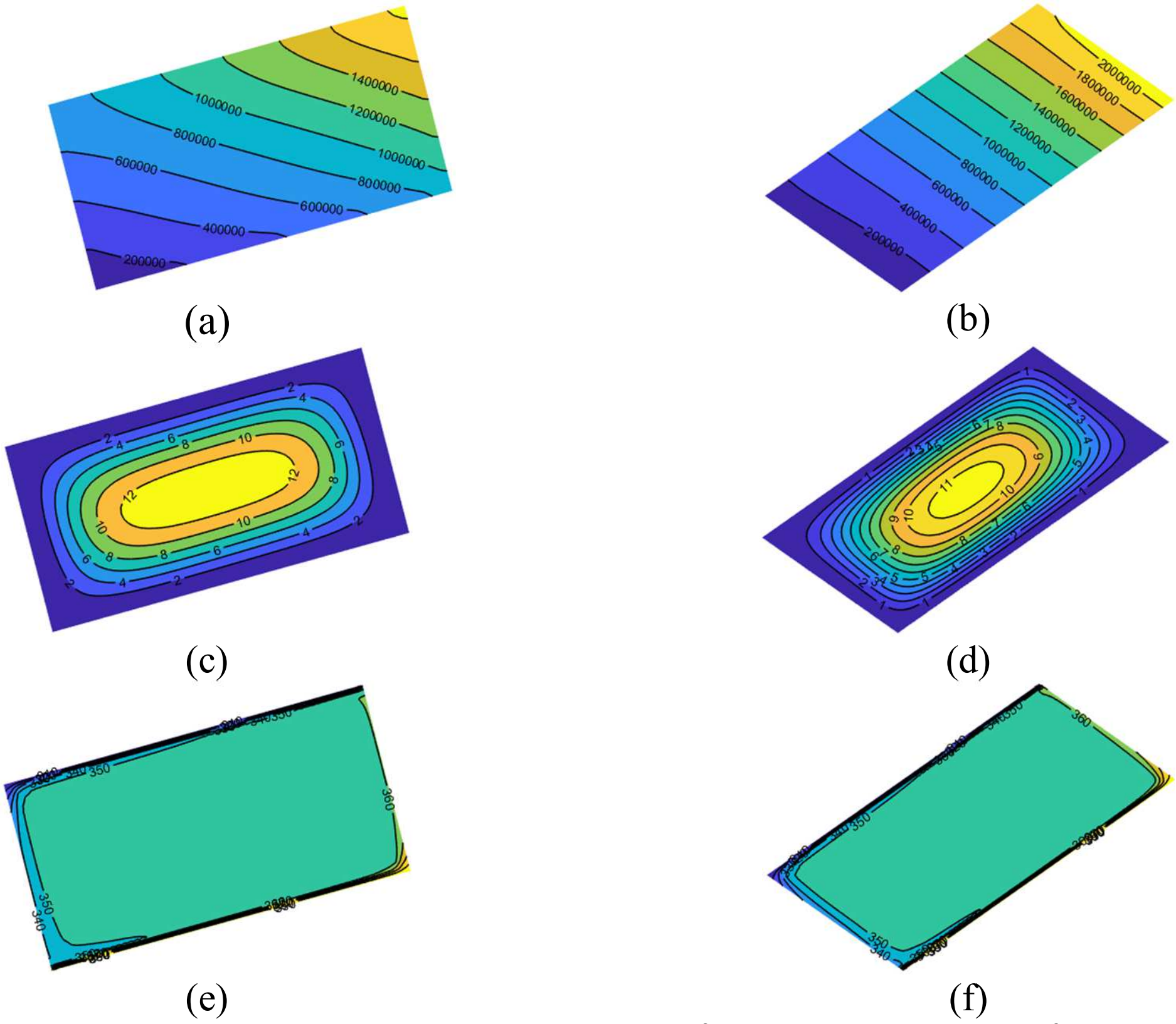

**Figure 6.** Pressure contours, streamlines, and isotherms for $\lambda = 15^0$ (left column) and $\lambda = 45^0$ (right column), $Ra = 10^5, Pr = 7.2, Le = 1,$ and $N = 1$.

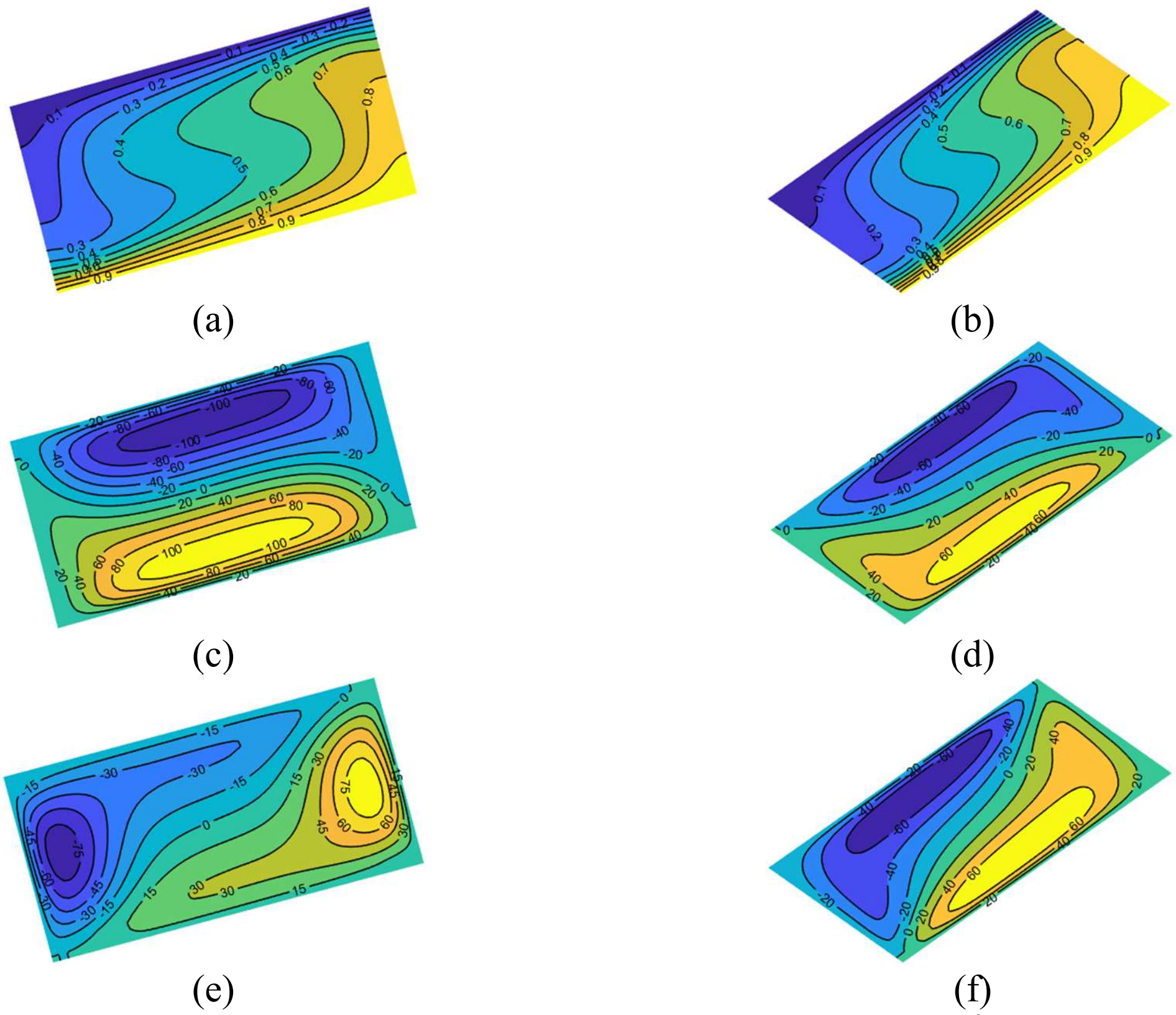

**Figure 7.** Iso-concentrations, $U$-velocity contours, and $V$-velocity contours for $\lambda = 15^0$ (left column) and $\lambda = 45^0$ (right column), $Ra = 10^5, Pr = 7.2, Le = 1,$ and $N = 1$.

From Figures 8(a) and (b), it is observed that pressure is low at the bottom left corner and high at the top right corner of the domain for $\lambda = 15^0$ and $\lambda = 45^0$. When the angle of inclination increases from $15^0$ to $45^0$, the numerical values of pressure increase but the nature of the contours are same. From figure 8(c) there is an anti-clockwise circulation formed at the center with two secondary circulations, one left of the center and other right of the center of the domain for $\lambda = 15^0$, and the intensity of the streamlines are strong near the center and weaker near the boundaries of the domain. From figure 8(d) there is an anti-clockwise circulation formed at the center with two secondary circulations, one at the top of the center and other one at the bottom of the center of the domain for $\lambda = 45^0$, and the intensity of the streamlines is strong near the center and weaker near the boundaries of the domain. When the angle of inclination increases from $15^0$ to $45^0$, the intensity decreases at the center and the nature of the streamlines are different. From figures 8(e) and (f) we can see that from the bottom to top the temperature decreases and the same thing happens from left to right. When the angle of inclination increases from $15^0$ to $45^0$, similar patterns are found for isotherms.

From Figures 9(a) and (b), the iso-concentration contours show that concentration decreases from the bottom to top. When the angle of inclination increases from $15^0$ to $45^0$, iso-concentrations almost follow the same pattern. From figure 9(c) we observed that there are two circulations, one is anti-clockwise and another one, clockwise, are formed in the domain for $\lambda = 15^0$. The anti-clockwise one is formed at the bottom, and the intensity is high near the center of the circulation; the clockwise one is formed at the top, and the intensity is weaker near the center of the circulation. From figure 9(d) we observed that there are two circulations; one is anti-clockwise and another one, clockwise, are formed in the domain for $\lambda = 45^0$. The anti-clockwise one is formed at the bottom, and the intensity is high near the center of the circulation; the clockwise one is formed at the top, and the intensity is weaker near the center of the circulation. When the angle of inclination increases from $\lambda = 15^0$ to $45^0$, the intensity decreases at the center of the bottom circulation and increases at the center of the top circulation. From figures 9(e) and (f) we observed that there are two circulations, one is anti-clockwise and other one, clockwise, are formed in the domain for $\lambda = 15^0$ and $\lambda = 45^0$. The anti-clockwise one is formed near the right wall, and the intensity is high near the center of the circulation; the clockwise one is formed near the left wall, and the intensity is weaker near the center of the circulation for $\lambda = 15^0$ . The anti-clockwise one is formed near the bottom wall, and the intensity is high near the center of the circulation; the clockwise one is formed near the top wall, and the intensity is weaker near the center of the circulation for $\lambda = 45^0$.

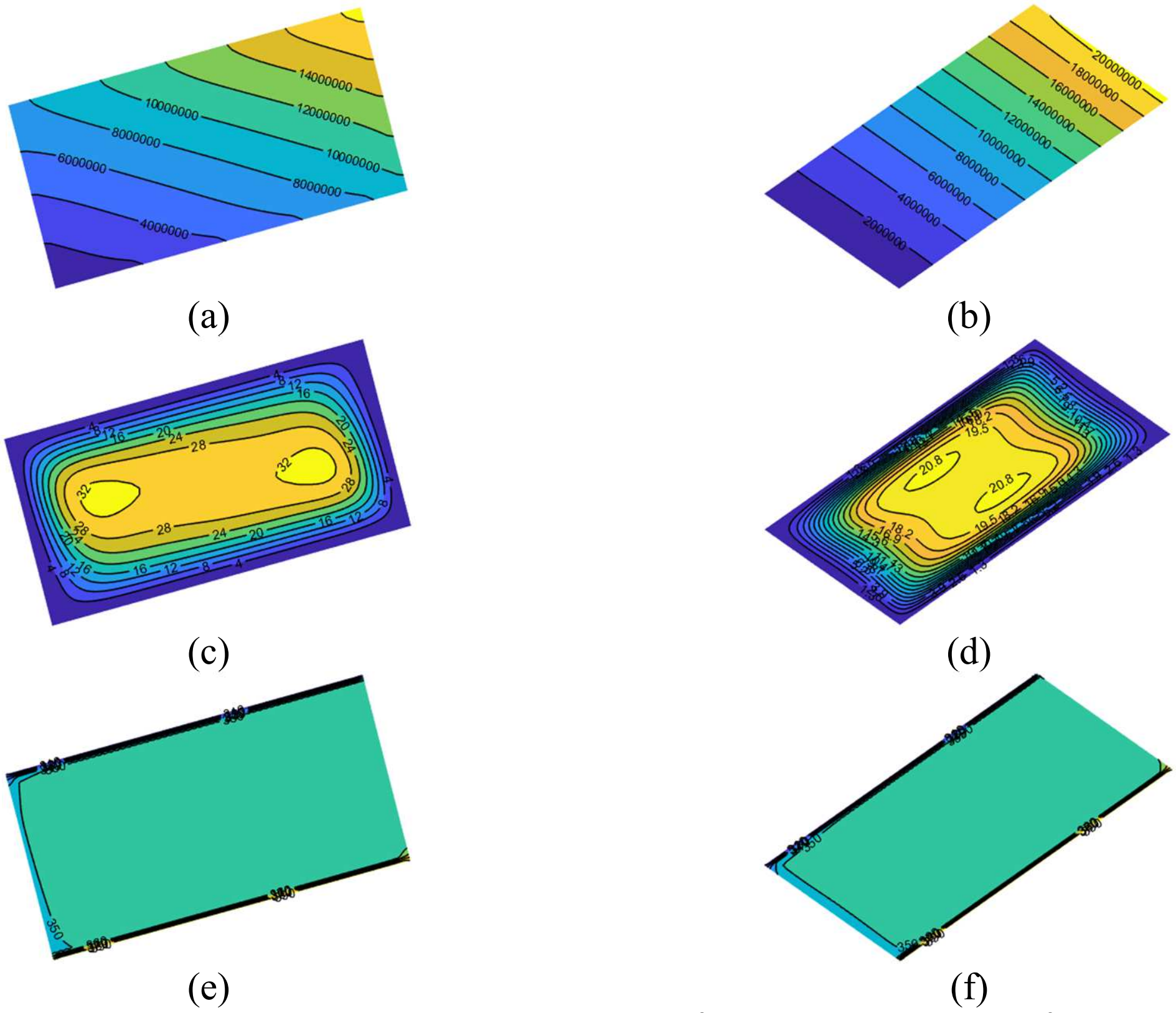

**Figure 8.** Pressure contours, streamlines, and isotherms for $\lambda = 15^0$ (left column) and $\lambda = 45^0$ (right column), $Ra = 10^6, Pr = 7.2, Le = 1,$ and $N = 1$.

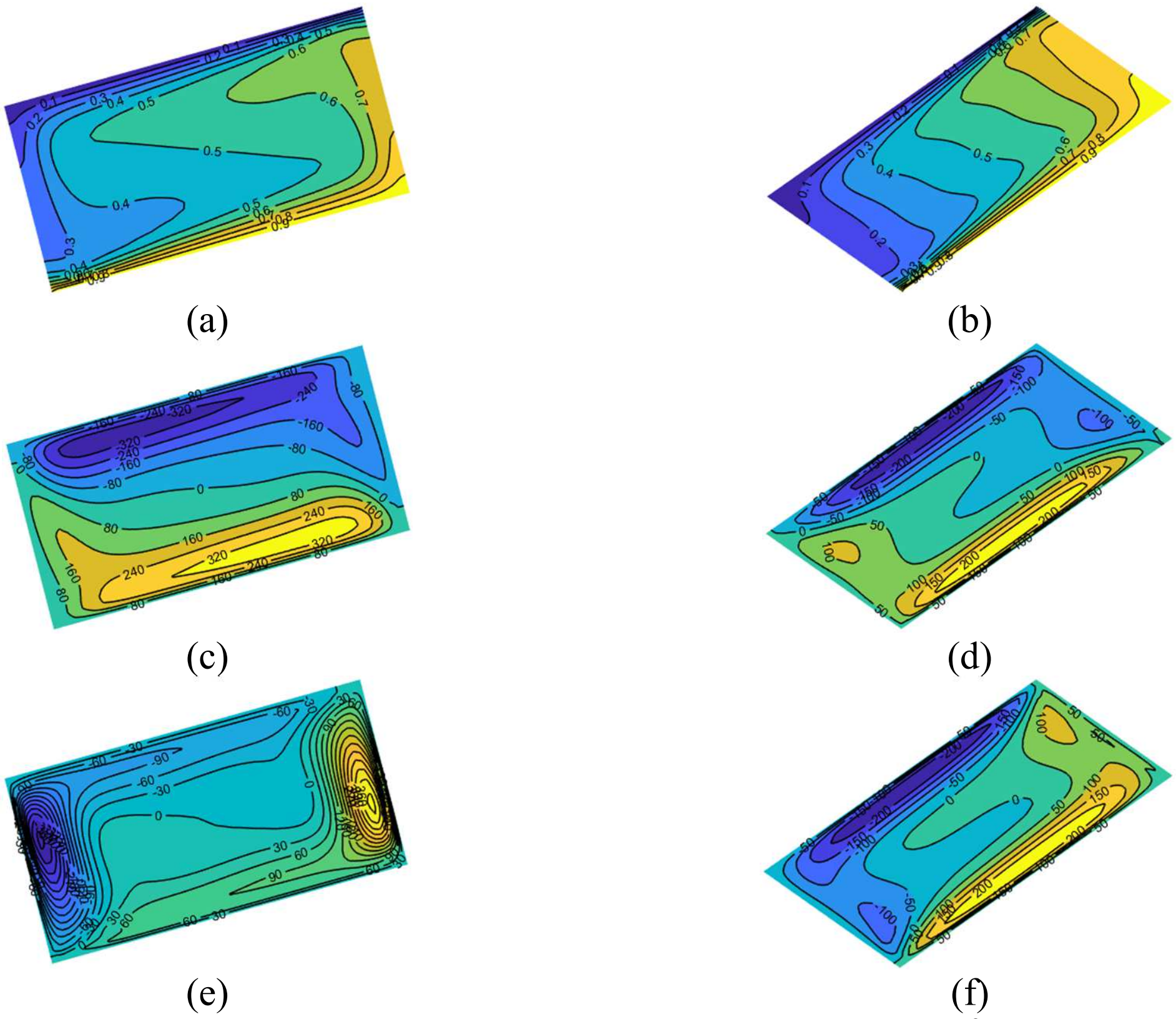

**Figure 9.** Iso-concentrations, $U$-velocity contours, and $V$-velocity contours for $\lambda = 15^0$ (left column) and $\lambda = 45^0$ (right column), $Ra = 10^6, Pr = 7.2, Le = 1,$ and $N = 1$.

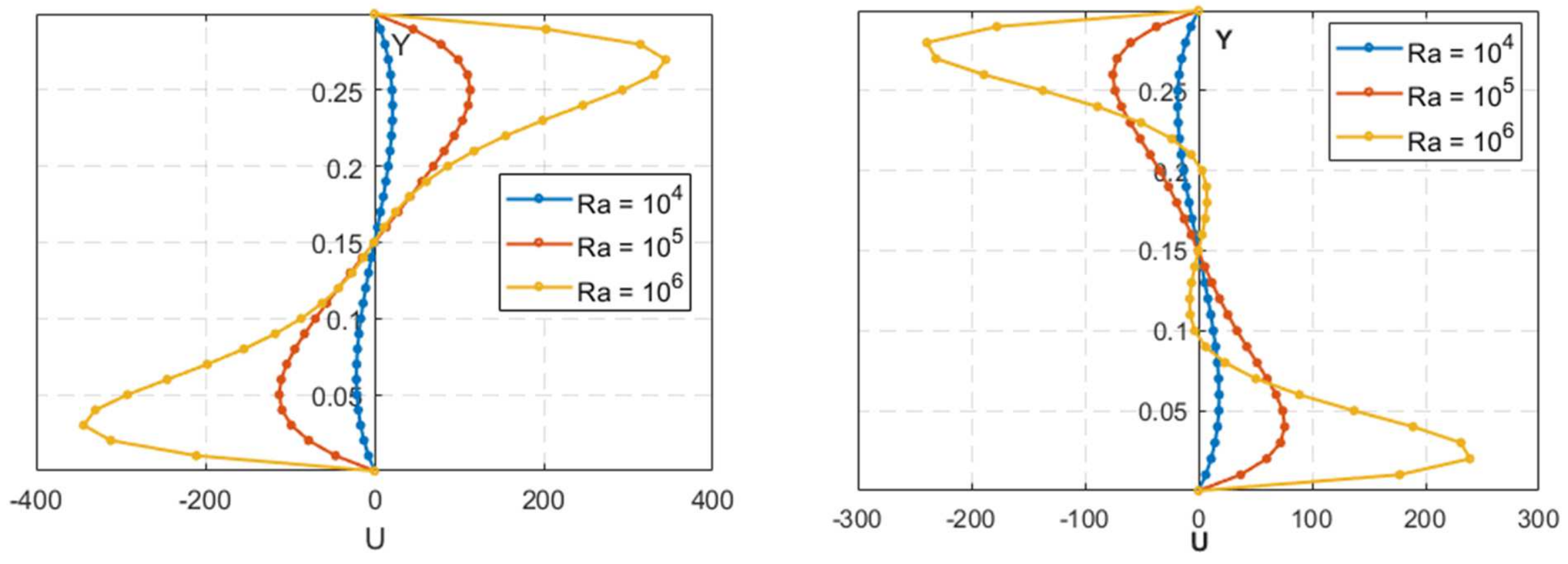

**Figure 10.** $U$-velocity along the vertical line through the geometric center of the domain $\lambda = 15^0$ (left) and $\lambda = 45^0$ (right), $Ra = 10^4 - 10^6, Pr = 7.2, Le = 1,$ and $N = 1$.

Based on the numerical solutions for $U$-velocity, Figure 10 illustrates the variation of $U$-velocity along the vertical line through the geometric center of the rectangular domain at Rayleigh numbers $10^4, 10^5,$ and $10^6$ for $\lambda = 15^0$ and $\lambda = 45^0$. We can see that $U$-velocity is high near the boundaries in comparison to the center. When the Rayleigh number increases, the magnitude of the $U$-velocity also increases. When the angle of inclination increases from $\lambda = 15^0$ to $\lambda = 45^0$, the magnitude of the $U$-velocity decreases.

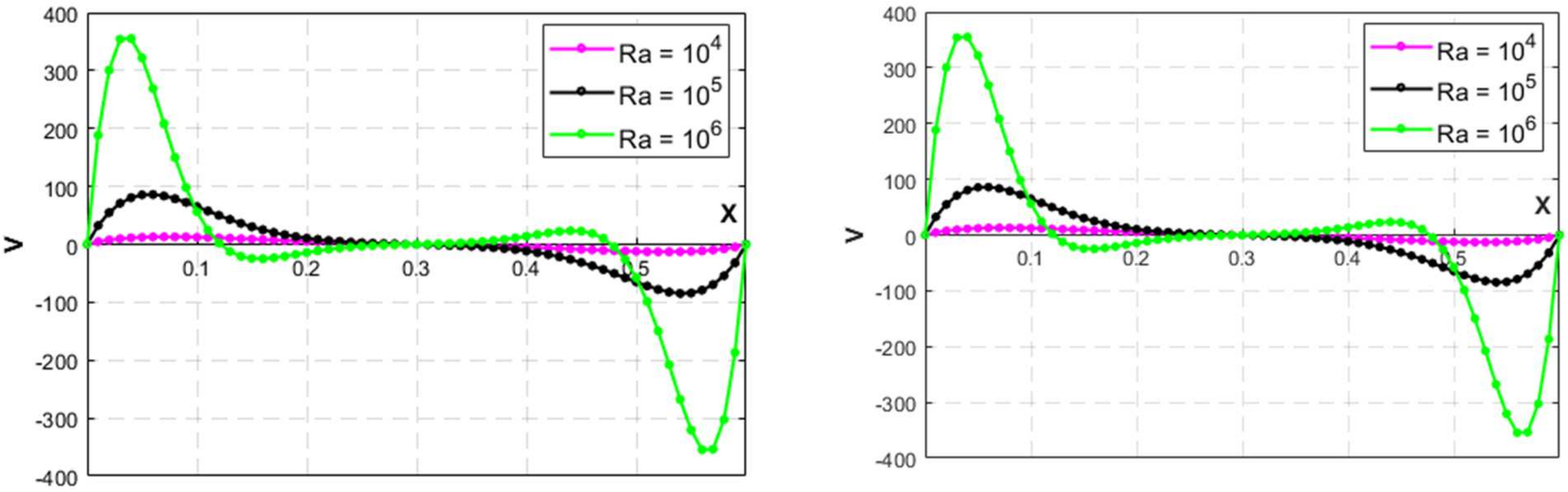

**Figure 11.** $V$-velocity along the horizontal line through the geometric center of the domain $\lambda = 15^0$ (left) and $\lambda = 45^0$ (right), $Ra = 10^4 - 10^6, Pr = 7.2, Le = 1,$ and $N = 1$.

Based on the numerical solutions for $V$-velocity, Figure 11 illustrates the variation of $V$-velocity along the horizontal line through the geometric center of the rectangular domain at Rayleigh numbers $10^4, 10^5,$ and $10^6$. We can see that $V$-velocity is high near the boundaries in comparison to the center. When the Rayleigh number increases, the magnitude of the $V$-velocity also increases. When the angle of inclination increases from $\lambda = 15^0$ to $\lambda = 45^0$, the magnitude of the $V$-velocity doesn't vary.

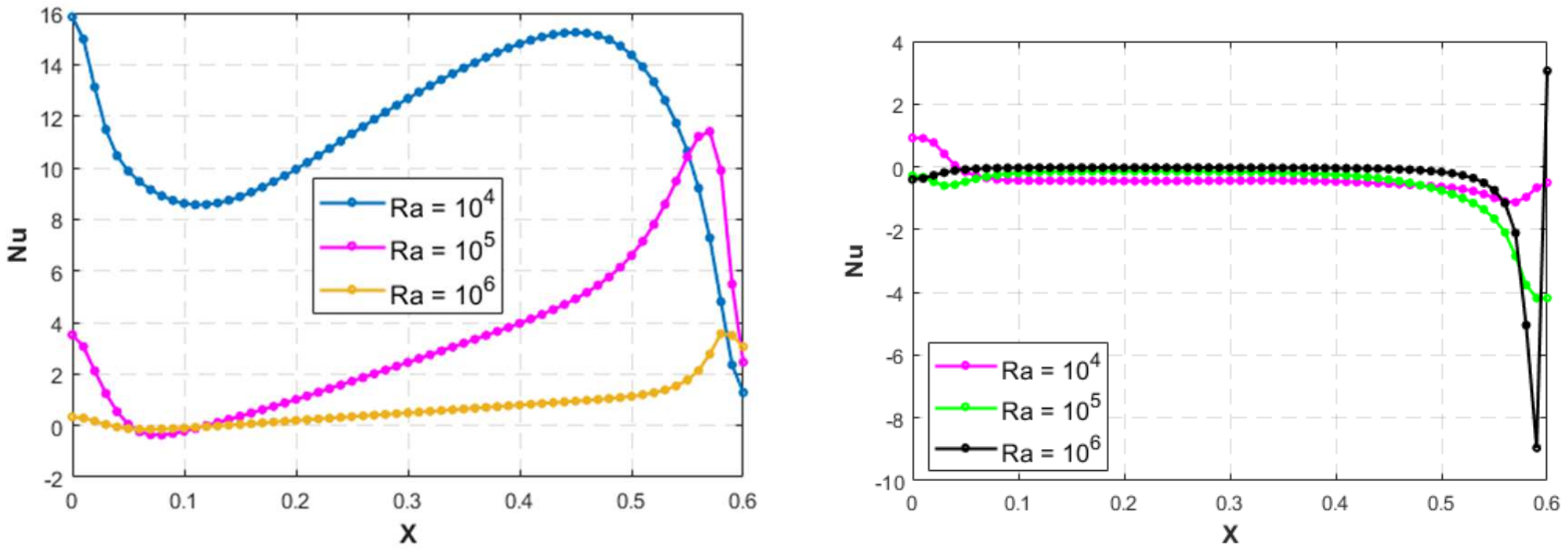

**Figure 12.** Local Nusselt number along the cold wall for $\lambda = 15^0$ (left) and $\lambda = 45^0$ (right), $Ra = 10^4 - 10^6, Pr = 7.2, Le = 1,$ and $N = 1$.

Based on the numerical solutions for the local Nusselt number, Figure 12 illustrates the variation of the local Nusselt number along the cold wall of the rectangular domain at Rayleigh numbers $10^4, 10^5,$ and $10^6$. We observed that the local Nusselt number decreases when the Rayleigh number increases. When the angle of inclination increases from $\lambda = 15^0$ to $\lambda = 45^0$, the local Nusselt number decreases.

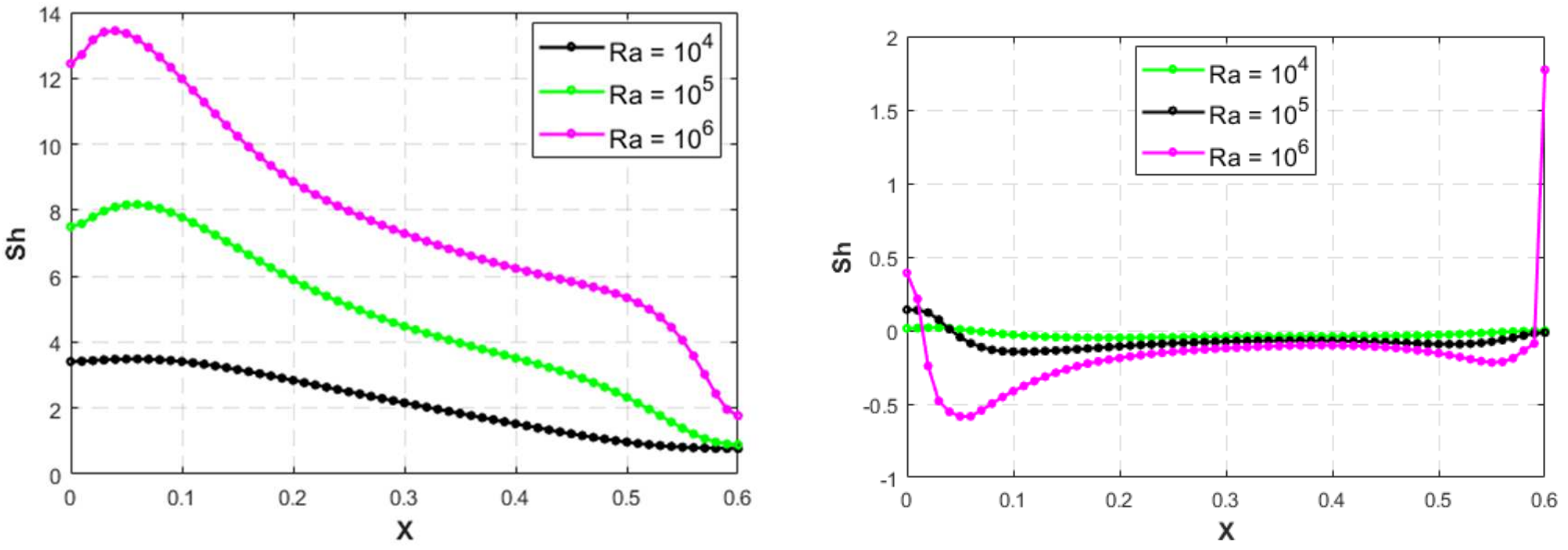

**Figure 13.** Local Sherwood number along the cold wall for $\lambda = 15^0$ (left) and $\lambda = 45^0$ (right), $Ra = 10^4 - 10^6, Pr = 7.2, Le = 1,$ and $N = 1$.

Based on the numerical solutions for the local Sherwood number, Figure 13 illustrates the variation of the local Sherwood number along the cold wall of the rectangular domain at Rayleigh numbers $10^4, 10^5,$ and $10^6$. We observed that the local Sherwood number increases when the Rayleigh number increases for $\lambda = 15^0$ and decreases for $\lambda = 45^0$.

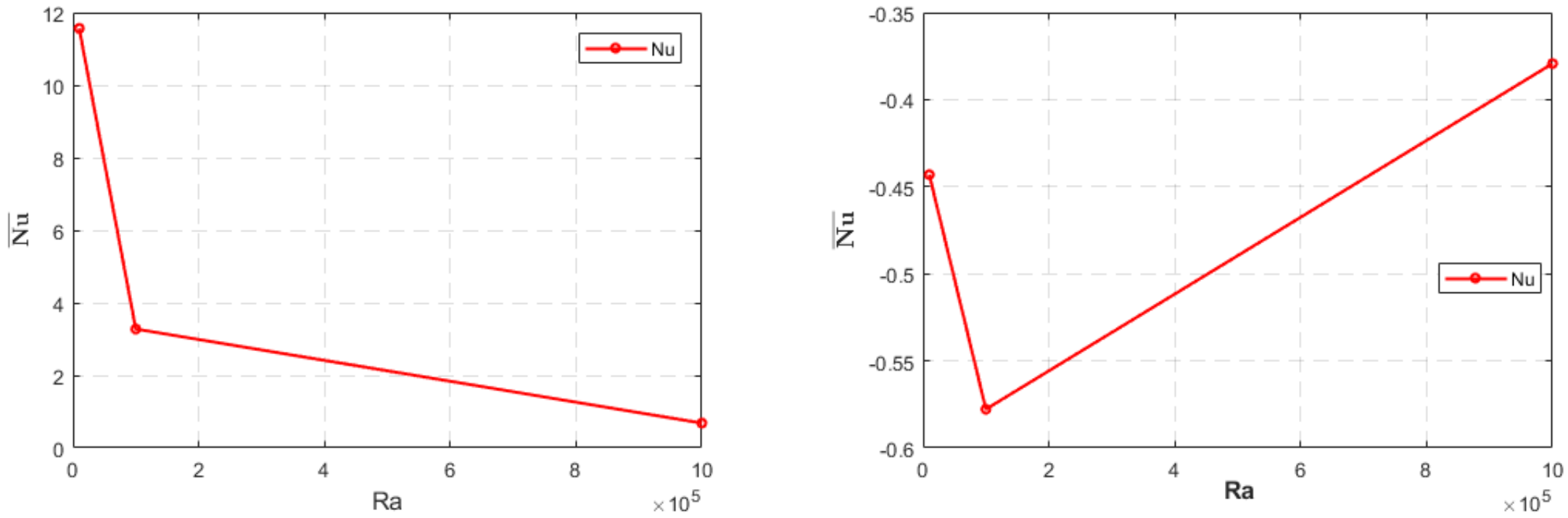

**Figure 14.** Average Nusselt number as a function of Rayleigh numbers for $\lambda = 15^0$ (left) and $\lambda = 45^0$ (right).

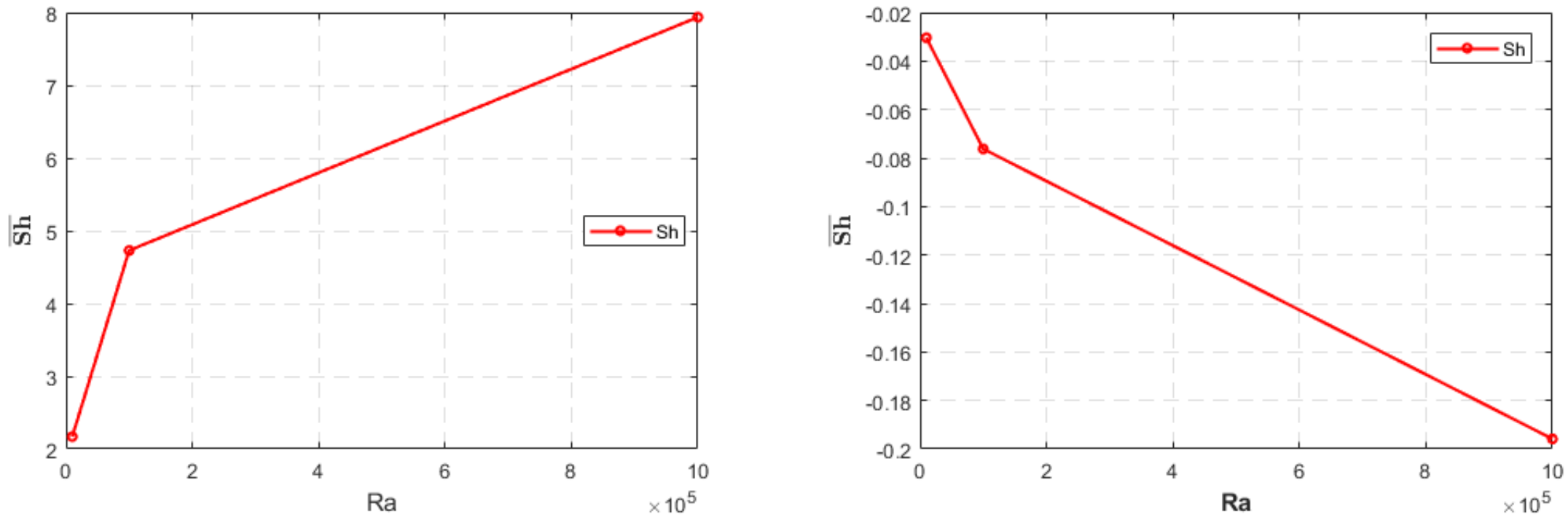

**Figure 15.** Average Sherwood number as a function of Rayleigh numbers for $\lambda = 15^0$ (left) and $\lambda = 45^0$ (right).

*Overall heat and mass transfer*: Based on the numerical solutions for the average Nusselt and Sherwood numbers, Figures 14 and 15 illustrate the variation of the average Nusselt and Sherwood numbers as a function of Rayleigh numbers. We can see that, as $Ra$ is increased from $10^4$ to $10^6$ the average Nusselt number is decreased for $\lambda = 15^0$, and decreases and then increases for $\lambda = 45^0$, whereas, the average Sherwood number increases for $\lambda = 15^0$ and decreases for $\lambda = 45^0$.

## 5. CONCLUSIONS

This paper presents the QUICK scheme of the finite volume method to investigate the problem of steady 2-D free convective incompressible flow with heat and mass transfer in an inclined rectangular domain at different Rayleigh numbers in the range of $10^4 \leq Ra \leq 10^6$, for Prandtl number $Pr = 7.2$, and Lewis number $Le = 1$. With this scheme, we discretize the governing equations along with the boundary conditions chosen in the present problem. The SIMPLE algorithm is adapted to compute the numerical solutions of the flow variables, local and average Nusselt and Sherwood numbers at different Rayleigh numbers as the range mentioned above. Our numerical solutions for $U$-velocity along the vertical line through the geometric center has been compared with benchmark solutions available in the literature for $\lambda = 15^0,\ Ra = 10^4$, $Pr = 7.2,$ and $Le = 1$ and it has been found that it is in the good agreement.

The magnitude of $U$-velocity along the vertical line through the geometric center of the rectangular domain increases with increasing Rayleigh numbers and decreases with increasing the angles of inclination. The magnitude of $V$-velocity along the horizontal line through the geometric center of the rectangular domain increases as the Rayleigh number is increased and doesn't vary with the increasing angles of inclination.

When the Rayleigh number and the angle of inclination increase, the value of pressure increases, but the nature of the contours remains the same. From streamlines, it is noted that when the Rayleigh number increases, the intensity increases near the center, and secondary cells are formed left and right of the center. When the angle of inclination increases, the intensity decreases near the center, and secondary cells are formed at the top and bottom of the center of the domain. This has happened because of the source terms in the momentum equations. From isotherms, it is noted that when the Rayleigh number and the angle of inclination increase, the temperature decreases from the bottom to top and the same thing happens from left to right.

When the Rayleigh number and the angle of inclination increase, the concentration decreases from top to bottom but the behavior changes. From the $U$-velocity contours, it is noted that when the Rayleigh number increases, the intensity increases in the bottom circulation and decreases in the top circulation. When the angle of inclination increases, the intensity decreases in the bottom circulation and increases in the top circulation. From the $V$-velocity contours, it is noted that when the Rayleigh number increases, the intensity increases in the bottom circulation and decreases in the top circulation, and the location of circulation changes as well. When the angle of inclination increases, the intensity decreases in the bottom circulation and increases in the top circulation, and the location of circulation changes also.

The local Nusselt number along the cold wall decreases when the Rayleigh number and angles of inclination increase. The local Sherwood number along the cold wall increases when the Rayleigh number increases for $\lambda = 15^0$ and decreases for $\lambda = 45^0$. We can see that, as the Rayleigh number increases, the average Nusselt number is decreased for $\lambda = 15^0$, and decreases and then increases for $\lambda = 45^0$ whereas the average Sherwood number increases for $\lambda = 15^0$ and decreases for $\lambda = 45^0$.

## NOMENCLATURE

| | |
|---|---|
| $\frac{\partial}{\partial n}$ | differentiation along the normal to the boundary |
| AR | aspect ratio, L/H |
| $p$ | pressure, Pa |
| $P$ | dimensionless pressure |
| $Nu_X$ | local Nusselt number |
| $\overline{Nu}$ | average Nusselt number |
| H | domain height, m |
| L | domain length, m |
| $T'$ | temperature, K |
| $T$ | dimensionless temperature |
| $x, y$ | cartesian coordinates |
| $X, Y$ | dimensionless Cartesian coordinates |
| $u, v$ | velocities components in $x$, $y$ direction, $m.s^{-1}$ |
| $U, V$ | dimensionless velocities |
| $Ra$ | Rayleigh number |
| $Pr$ | Prandtl number |
| $\overline{Sh}$ | average Sherwood number |
| $Sh_X$ | local Sherwood number |
| $C'$ | concentration |
| $C$ | dimensionless concentration |
| $Le$ | Lewis number |
| $N$ | buoyancy ratio |
| $g$ | gravitational acceleration, $m.s^{-2}$ |

**Greek symbols**

| | |
|---|---|
| $\alpha$ | thermal diffusivity, $m^2.s^{-1}$ |
| $\nu$ | kinematic viscosity, $m^2$.s-1 |
| $\Psi$ | stream function |
| $\rho$ | fluid density, $kg.m^{-3}$ |
| $\lambda$ | angle of inclination |
| $\mu$ | dynamic viscosity, kg. $m^{-1}.s^{-1}$ |
| $D$ | mass diffusivity, $m^2.s^{-1}$ |
| $\beta_{T'}$ | coefficient of thermal expansion, $K^{-1}$ |
| $\beta_{C'}$ | coefficient of concentration expansion, $m^3$.kg |

**Subscripts**

| | |
|---|---|
| $h$ | hot |
| $c$ | cold |

## ACKNOWLEDGMENT

The first author acknowledges the support from research council, the University of Delhi for providing research and development grand 2015–16 vide letter No. RC/2015-9677 to carry out this work.